\documentstyle [12pt] {article}
\newcommand{\cs}[3]{{{#3} \brace {#1 #2}}}
\newcommand{\h}[1]{\mathop{\lambda}\limits_{#1}\ \!\!\!}
\newcommand{\pder}[2]{\frac{\partial{#1}}{\partial{#2}}}

\begin{document}

\title{\bf Motion of Spinning Particles in Gravitational Fields}
\author{{\bf M.I.Wanas}\\
\normalsize {\it Astronomy Department, Faculty of Science,}\\
{\it Cairo University, Egypt}~~  e.mail: wanas@frcu.eun.eg}
\maketitle
\begin{abstract}

    A new path equation in absolute parallelism (AP) geometry is derived.  The 
equation is a generalization of three path equations derived in a previous work.
It can be considered as a geodesic equation modified by a torsion term, whose 
numerical coefficient jumps by steps of one half.  The torsion term is parametrized
using the fine structure constant.  It is suggested that the new equation may
describe the trajectories of spinning particles under the influence of a
gravitational field, and the torsion term represents a type of interaction
between the quantum spin of the moving particle and the background field.

    Weak field limits of the new path equation show that the gravitational potential felt by a spinning
particle is different from that felt by a spinless particle (or a macroscopic
body).

    As a byproduct, and in order to derive the new path equation, the AP-space
is reconstructed using a new affine connexion preserving metricity.  The new
AP-structure has non-vanishing curvature.  In certain limits, the new AP-structure
can be reduced either to the ordinary Riemannian space, or to the conventional
AP-space.   

\end{abstract}
\vskip 1truecm
\baselineskip= .8cm
\section{Introduction}

It is well known that most of the astronomical informations are carried by
massless spinning particles.  Astronomers usually extract astronomical 
informations from photons, which are spin-1 massless particles.  In 1987
astronomers succeeded in extracting informations carried by nutrinos (sipn-
one half massless particles) coming from SN1987A.  In the near future we
expect gravitons (spin-2 massless particles) to open a third window through
which we look at the universe.  These particles, during their trip from their
sources to the detectors, are moving in different gravitational fields.  Thus,
it is of fundamental importance, for astronomy and space studies, to know the 
exact trajectories of such particles, and the interaction between the spin
of these particles and the background gravitational field.

    The problem of motion of spinning \footnote{In the present work we are going to distinguish between spin and rotation.
We use the term spin for the quantum property of a microscopic particle, while
the term rotation is used for the corresponding classical mechanical property
of a macroscopic object.} \ particles in a background gravitational
field has been tackled by different authors.  The trails of those authors can
be classified into two main approaches.  {\it The first approach} contains two
classes: (i) quantization of a relativistic equation of motion of a 
rotating object (e.g.Papapetrou equation, Dixon equation),(cf. Melek (1988))
(ii) geometrization (or more specifically coveriantization) of a quantum mechanical equation of
motion of a spinning particle (e.g. Schr\"{o}denger equation, Pauli equation),(cf. DeWitt (1957),
Galvao and Teitelboim (1980)).
{\it The second approach} depends on a different philosophy in which paths (or curves)
of an appropriate geometry are considerd to represent trajectories of test 
particles . Einstein had followed this approach and used the geodesic and null geodesic curves of the
Riemannian geometry to describe  the motion of test particles and of photons
respectively.

    Although the philosophy of the second approach is successful in describing motion
of test particles including massless particles (photons), yet it was neglected by many
authors in dealing with motion of other spinning particles. We believe that this philosophy      
deserves further investigations especially to look for the equation of motion of spinning 
particles  in  gravitational fields. It is to be noted that while the first approach 
is suitable for describing
short range motion of spinning particles i.e. motion on the laboratory scale, 
it is not suitable for describing long range motion (e.g. motion on scales such as solar 
system, galactic, intergalactic scales).  The second approach is more suitable
for describing trajectories of test particles especially  long range motion of massless particles in gravitational fields
(motion of photons using null-geodesics).  We are mainly interested in
this approach. 

To explore the capabilities of this approach we directed our attention to 
the AP-geometry. The cause is that  short range motion of spinning particles 
is described successfully using this  geometry, since spinors can be defined and related to
the structure of the AP-spaces.
   
    In a trial to look for possible paths in this geometry, three path equations
were derived (Wanas et. al. (1995)a) by generalizing the method given by 
Bazanski (1977,1989). These equations can be written in the form, 

$$
{\frac{dV^\mu}{dS^+}} + \{^{\mu}_{\alpha\beta}\} V^\alpha V^\beta = - 
\Lambda^{~ ~ ~ ~ \mu}_{(\alpha \beta) .} ~~~V^\alpha V^\beta,  \eqno{(1.1)} 
$$
$$
{\frac{dW^\mu}{dS^0}} + \{^{\mu}_{\alpha\beta}\} W^\alpha W^\beta = - {\frac{1}{2}} 
\Lambda^{~ ~ ~ ~ \mu}_{(\alpha \beta) .}~~~ W^\alpha W^\beta,  \eqno{(1.2)} 
$$
$$
{\frac{dU^\mu}{dS^-}} + \{^{\mu}_{\alpha\beta}\} U^\alpha U^\beta = 0,  \eqno{(1.3)}
$$           
where$S^{+}, S^{0}$ and $S^{-}$ are the evolution  parameters characterizing
the three paths  respectively ; and  $V^{\alpha}, W^{\alpha}$ and$U^{\alpha}$ are
the tangents to the corresponding paths and the brakets () are used for
symmetrization. The torsion of the AP-space is defiend by 
$$
\Lambda^\alpha_{. \mu \nu}{\ } {\stackrel{def.}{=}} {\ }  \Gamma^\alpha_{. \mu \nu}  - 
\Gamma^\alpha_{. \nu \mu} {\ }{\ },
$$
where  $\Gamma^{\alpha}_{.\mu\nu}$ is a non-symmetric affine connexion defiend as a consequance 
of the AP-condition (see section.2).
These equations can be considered as generalization of the geodesic equation
of Riemannian geometry.

Considering these three equations, it seems interesting
to point out the following remarks:

1- Although four absolute derivatives are defiend in the AP-space (see section 2), and used in 
deriving the above equations, only three equations were obtained inculding the 
geodesic equation (1.3).

2- In general, the other two equation ((1.1),(1.2)) can not be reduced to the 
geodesic  equation (1.3) unless the torsion vanishes . It has been shown
(Wanas,and Melek (1995) ) that the vanishing of the torsion of the AP-space will lead
to the vanishing of the curvature tensors defined in that space. In this
case the space will reduce to a flat one.

3- Equation (1.3) can be reduced to null-geodesic upon reparametrization.

4- The equations can be considered as geodesic equations modified by a torsion 
term. The important feature of this set of equations is that the numerical coefficient
of the torsion term jumps by a step of one half from one equation to the next.
This is tempting to beleive that  paths in the AP- spaces are quantized in
a certain sense. 

From these remarks it seems that the AP-space possesses more finer structure
than that appearing in its  conventional picture (for more details about the 
conventional sturture of the AP-spaces, (cf. Levi-Civita (1950), Mikhail (1952), Hayashi
and Shirafuji (1979)). It contains in addition to the geodesic and null- geodesic 
equations, two more equations (1.1), (1.2).

Now the following groups of questions emerge:

1- Does the space possesses  a general affine connexion that gives rise to 
family of paths in which the coefficient of the torsion term jumps by a step
of $\frac{1}{2}$ from equation to another in this family ? If so, what is the underlying geometric
structure ? 

2- What is the general form of the equations representing this family with a 
jumping torsion term ?

3- Is there any observational or experimental evidence for motion alonge such paths ?  What is the order of 
magnitude of the deviation from the geodesic motion, if any? What is the 
intrinsic property of the moving particle that causes such deviation from the
geodesic one?

    The aim of the present work is to give answers to some of the above questions.
In section 2, a trail to answer the first group of the above questions is given.  In section 
3 the general form of a path equation, giving the required family mentioned
above, is derived using the geometry  given in section 2.  Section 4 contains
a trail to attribute some physical meaning to the new path equation.  The static 
weak field limits of the new equation are given in section 5.  The work is discussed in
section 6.

\section{The Underlying Geometry}  
                                          
    In the conventional AP-geometry two affine connexions  were defined.  The first is
Christoffel symbol defined as a consequence of the metricity condition,
$$
g_{\alpha \beta ; \mu} = 0,  \eqno{(2.1)}
$$     
where $g_{\alpha \beta}$ is the metric tensor defiend by, 
$$
g_{\alpha \beta} {\ } {\stackrel{def.}{=}} {\ } \h{i}_{\alpha} \h{i}_{\beta} {\ }{\ }{\ },  \eqno{(2.2)}
$$ 
and $\h{i}^{\alpha}$ are the tetrad vectors defining the structure of the AP-space 
in 4-dimensions. The semicolon is used to characterize covarient dervatives using
Christoffel symbols. The second is the non-symmetric connexion 
${\Gamma}^{\mu}_{.\alpha \beta}$ defiend as a consequence of
absolute parallelism condition:
$$
{{\h{i}}^{{\stackrel{\alpha}{+}}}_{~~.| \beta}} = 0 {\ }{\ }{\ },    \eqno{(2.3)}
$$
the stroke and the (+) sign is used to characterize absolute derivatives using
${\Gamma}^{\mu}_{.\alpha \beta}$. Since this connexion is non-symmetric, one 
can define two more absolute derivatives : for the first we use a stroke 
and a (-) sign which indicates the use of the dual connexion 
 ${\hat{\Gamma}}^{\mu}_{.\alpha \beta}  (= {\Gamma}^{\mu}_{.\beta \alpha})$, and 
for the second we use a stroke without signs, which indicates the use
of the symmetric part,  $\Gamma^{\mu}_{.(\alpha \beta)}$. It can be shown
that using (2.2), (2.3) we get, 
$$
g_{{\stackrel{\mu}{+}}{\stackrel{\nu}{+}}| \sigma} =0  {\ }{\ }{\ },             \eqno{(2.4)}
$$        
which means that metricity is also preserved using the non-symmetric 
connexion, recalling that the metricity condition (2.1) is necessary but not sufficient
to define Cristoffel symbol. 

Now we are looking for a general affine connexion that is capable of producing
the family of all paths with jumping torison term .Recalling that the three 
path equations were derived using the above mentioned derivatives ,thus      
the simplest way to find such a connexion is to combine linearly the above connexion using
some parameters.After some manipulations the general expression of this connexion can be written 
in the following form,  
$$
\nabla^\mu_{.\alpha \beta} = a_1 \{^\mu_{\alpha \beta}\} + (a_2 - a_3) 
\Gamma^{\mu}_{. \alpha \beta} - (a_3 + a_4)\Lambda^{\mu}_{. \alpha \beta}{\ }{\ }{\ },               
                                                                  \eqno{(2.5)}
$$
where $a_1 , a_2, a_3$ and $a_4$ are parameters to be fixed later.
It can be easily shown that $\nabla^{\alpha}_{.\mu \nu}$ transforms as an affine
connexion under the group of general coordinate transformations. It is clear 
that this connexion is non-symmetric. If we charactize the general absolute 
derivative, using the affine connexion (2.5), by a double stroke we get after
some manipulations:    
$$
g_{\mu \nu || \sigma} = (1- a_1 - a_2 - a_3 -2a_4)(g_{\mu \alpha} \{^{\alpha}_{\sigma\nu}\}
+ g_{\nu \alpha} \{^{\alpha}_{\mu\sigma}\}) - (a_3 + a_4)(g_{\mu \alpha}\gamma
^{\alpha}_{. \sigma \nu} + g_{\nu \alpha}\gamma^{\alpha}_{. \sigma\mu}){\ },
                                                              \eqno{(2.6)}
$$
where  ${\gamma}^{\alpha}_ {.\mu \nu} $ is the contorsion defined by 
$$
\gamma^{\alpha}_{. \mu \nu} {\ } {\stackrel{def.}{=}} {\ }   \frac{1}{2}(\Lambda^{\alpha}_{.\mu \nu} -
{\Lambda}_{\mu .\nu }^{~\alpha} - {\Lambda}_{\nu .\mu}^{~\alpha})
                                                                \eqno{(2.7)}
$$
If we need metricity to be prserved along the paths characterized by (2.5)
we should take,

$$
a_3+a_4=0.
$$
$$ 
 1 - a_1 - a_2 + a_3 = 0, 
$$
Taking $a=a_1$, $b=a_2+a_4$ then the metricity condition can be written in the
form

$$
  a+b=1  .   \eqno{(2.8)}
$$
After using this condition it is clear that (2.5) will remain non-symmetric.

Using the general affine connexion (2.5) and the general metricity condition 
(2.8), one can define the following curvature tensors.

(i) If we replace, in the definition of Riemann-Christoffel tensor    
$R^{\alpha}_{.\mu \nu \sigma}$, Christoffel symbol by the new connexion
we get the curvature,
$$
B^{\alpha}_{.\mu \nu \sigma} {\ }  {\stackrel{def.}{=}} {\ }  \nabla^{\alpha}_{.\mu \sigma ,\nu} -
\nabla^{\alpha}_{.\mu \nu ,\sigma} + \nabla^{\alpha}_{.\epsilon  \nu}
\nabla^{\epsilon}_{.\mu \sigma } - \nabla^{\alpha}_{.\epsilon  \sigma }
 \nabla^{\epsilon}_{.\mu  \nu }{\ }{\ }{\ },       \eqno{(2.9)a}
$$ 
which can be written in the form  
$$
B^{\alpha}_{.\mu \nu \sigma} = R^{\alpha}_{.\mu \nu \sigma} + 
b {\gamma}^{\alpha}_ {.\mu [\sigma ; \nu]} +b^2 {\gamma}^{\alpha}_ {.\epsilon [\nu}
{\gamma}^{\epsilon}_ {.\mu \sigma ]} {\ }{\ }{\ },               \eqno{(2.9)b}
$$
the square brackets are used for antisymmetrization.

(ii) If we define the curvature $W^\alpha_{.\mu \nu \sigma}$ as a measure 
 of the non-commutation of the  general absolute derivatives, we get
$$
\h{i}_{\mu || \nu \sigma} - \h{i}_{\mu || \sigma \nu}  =  \h{i}_{\alpha}{\ }W^{\alpha}_{. 
\mu \nu \sigma}                                              \eqno{(2.10)}
$$
where 
$$
W^\alpha_{.\mu \nu \sigma} = B^\alpha _{.\mu \nu \sigma} - b(b-1) \gamma
^\alpha_{.\mu \epsilon} \Lambda^\epsilon_{.\nu \sigma}  \eqno{(2.11)}
$$

Now we have the following remarks:  (1) It is to be considered that non of the 
curvature tensors defined by (2.9) and (2.11) vanishes. This means 
that the new structure of the AP-space is, in general not flat.  This result
will be discussed in section 6.   (2) Taking the metricity condition into 
consideration, we can show that two important special cases could be obtained:

{\it The first case} a=1 (i.e. b=0): In this case the geometry will be reduced  to 
Riemannian geometry.  This is obvious by substituting the values of the parameters
into (2.5), (2.9)b, and (2.11) , which give ,
$$
\nabla^\alpha _{.\mu \nu} =  \{^{\alpha}_{\mu \nu}\}  \eqno{(2.12)}
$$
$$
W^\alpha _{.\mu \nu \sigma} =  B^\alpha_{.\mu \nu \sigma} = R^\alpha _{.\mu \nu \sigma}     \eqno{(2.13)}
$$ 

{\it The second case} a=0 (i.e. b=1): In this case it can be easily shown, after
some manipulations, that the geometry reduces to the conventional AP-geometry with
$$   
\nabla^{\alpha}_{.\mu \nu} = \Gamma^{\alpha}_{.\mu \nu}   \eqno{(2.14)}
$$
$$
B^{\alpha}_{.\mu \nu \sigma} {\ }  = {\ } \Gamma^{\alpha}_{.\mu  \sigma, \nu} - 
\Gamma^{\alpha}_{.\mu \nu, \sigma} + \Gamma^{\alpha}_{\epsilon \nu} \Gamma^
{\epsilon}_{. \mu \sigma} - \Gamma^{\alpha}_{. \epsilon \sigma} 
\Gamma^{\epsilon}_{. \mu \nu}  =    0     \eqno{(2.15)}
$$
$$
W^{\alpha}_{.\mu \nu \sigma}  =  0.    \eqno{(2.16)}
$$
The curvature  (2. 15) vanishes as a consequence of the
AP-condition.

These two cases will be discussed in the last section. Now to complete the   
strcture of the space, one should look for a path  equation correspoding
to the new affine connexion, which  can be considered as a generalization
of equations (1.1), (1.2) and (1.3). This is done in the following section.

\section{The  General Form of the Path Equation}

Generalizing the Bazanski (1977,1989) Lagrangian, using the affine connexion (2.5)
and the condition (2.8), we get 
$$
L{\ }{\stackrel{def.}{=}}{\ } \h{i}_{\mu} \h{i}_{\nu} Z^{\mu}{{\frac{\nabla{\chi}^\mu}{ \nabla\tau}}}
                 \eqno{(3.1)}
$$
$$
Z^{\mu} {\ } {\stackrel{def.}{=}} {\ }  {\frac{dx^\mu}{d\tau}},        \eqno{(3.2)}
$$
$$
{\frac{\nabla {\chi}^\mu}{ \nabla\tau}} {\ } {\stackrel{def.}{=}} {\ } \chi^{\mu}_{.|| \alpha}Z^{\alpha}
                                           \eqno{(3.3)}
$$
$$
\chi^{\mu}_{.|| \alpha} {\  } {\stackrel{def.}{=}} {\  } \chi^{\mu}_{., \alpha} + \chi^{\beta}\nabla^{\mu}_{. \beta \alpha}  ,
                                      \eqno{(3.4)}
$$
where $\tau$ is the evolution parameter along the new general path associated with (2.5), $\chi^{\mu}$ is
the deviation vector, and $Z^{\mu}$ is the tangent to the path. Applying the variational formalism using the lagrangian (3.1), and noting that the general absolute
differentional operation commmutes with rasing and lowering indices (because
of (2.8)), we get after necessary manipulation :
$$
{{\frac{dZ^\mu}{d\tau}} + \cs{\nu}{\sigma}{\mu}\ Z^\nu Z^\sigma = 
- b{\ } \Lambda_{(\nu \sigma)}.^\mu~~  Z^\nu Z^\sigma}  .\eqno{(3.5)a}
$$    
Using (3.3), the last equation can be written in the form
$$
{\frac{\nabla Z^\mu}{ \nabla\tau}} = 0  .  \eqno{(3.5)b}
$$ 
Equation (3.5)a, or (3.5)b, represents a generalization of the path equations  
given in section 1. Moreover, if $b=0$ this equation will reduce to geodesic of the metric 
(or null-geodesic upon reparametrization). This is consistent with the first
special case given in the above section. It can be easily shown that (3.5) will
give rise to,
$$
\frac{d(g_{\alpha \beta}Z^\alpha Z^\beta)}{d\tau} =  0
$$               
consequently, its first integral is given by, 

$$			
Z^\alpha{\ }Z_{\alpha} =  Z^2                       \eqno{(3.6)}
$$
which means that Z is a constant along the path (3.5), and since Z is scalar under 
general coordinate transformation, then it will be constant in general.   

\section{Physical Meaning of the Torsion Term} 
    
Equation (3.5) reprsents a new path in the AP-space. As stated above, this equation
can be reduced to a geodesic equation (as b=0) which represents the trajectory
of a test particle in a background gravitational field (it can also be reduced
to a
null-geodesic). So, what is the  role of the torsion term  on the R.H.S. of (3.5)a ?
May it repesent a type of interaction between the torsion of the background 
gravitational field and some intrinsic  property of the moving particle ?  	
If so, what is this intrinsic property ? 

Let us start a trial to answer the last question. Recalling that, the aim of the work
in the section 3 is to get an equation that reprsents the complete family of
the equations given in section 1. Thus the parameter "b" of (3.5)a should consist
of a half integer part. In this case we can write 
$$
b = {\frac{n}{2}}{\ }\beta ,                          \eqno{(4.1)}
$$      
where n is a natural number  and $\beta$ is another parameter.

It is well known that an intrinsic property 
of the particle which depends on half integers is its quantum spin. 
Several authors have pointed out that spinning particles feel space-time torsion
(cf. Hehl (1971)). Others beleive that spinning particles are the only probes which
detect telleparallel geometry(cf. Nitch and Hehl (1980), Ross (1989)). Although
most of these authors have used the term spin to mean rotation and have believed that only
rotating sources can generate the space-time torsion, the situation here is different.
It has been shown by the author(1990) that space-time torsion 
is generated whether the source of
the field is rotating or not.  Considering these arguments,  we may suggest that the R.H.S. of (3.5)a reprsents a type of interaction between 
the quantum spin of the moving particle and the torsion of the background gravitational
field. Consequently, equation (3.5) may reprsent the trajectory of a spinning 
particle in a gravitational field. Consequently we take $n = 0, 1, 2,...$ for
particles with spin 0 , ${\frac{1}{2}}$, 1,.....  respectively. For macroscopic 
objects and spinless particles $n = 0$.This will reduce equation (3.5)  to the geodesic 
(or null-geodesic) equation, and the geometric structure to a Riemannian one.

It is well known that the geodesic motion
implies the validity of the weak equivalence principle (WEP). So, equation (3.5)
implies that motion of spinning particles violates the WEP.
If we take the parameter $\beta$ to be of order unity, then the torsion term will
be of the same order of magnitude as the Christoffel symbol term which will be considerably
large. But since there are no experimental or observational evidences for such large violation of (WEP)
for the motion of spinning particles, thus the parameter $\beta$ should be of  
less order. 
From observational point of view the WEP is varified to an accuracy of about $10^{-2}$ 
on the galactic scale (cf. Longo (1988)), so $\beta$ should be less than  $10^{-2}$.

From the previous discussion it is acceptable that the parameter $\beta$ 
may have the following properties :

(1) It should be a dimensionless quantity to preserve the dimensions on both sides
of (3.5)a.

(2) It should be small compared to unity $(\beta \leq 10^{-2})$   to be consistent with relevant
obersvations and experiments.

(3) It should be connected to the intrinsic properties of elementary particles, 
 especially  those affecting motion of the spinning particles.

(4) It should  include, in its structure, Plank's constant $\hbar$ or h.    
      
To the knewledge of the author a quantity  satisfying the above requirements is the 
the fine structure constant $\alpha (={\frac{e^2}{\hbar c}} ={\frac{1}{137}})$.
We can replace $\beta$ in (4.1) by the fine structure constant $\alpha$. But to 
be more conservative we are going to write (4.1) in the form   
$$
b = {\frac{n}{2}}{\alpha {\ } \gamma}             \eqno{(4.2)}
$$ 
where $\gamma$ is a dimensionless parameter of order unity to be fixed by experiment.
Now the parameter b constitutes of three parts: the first" ${\frac{n}{2}}$" is the part that
makes (3.5)a gives rise to (1.1), (1.2), (1.3), and the second part "$\alpha$" acts 
as a reduction factor, while the third part"${\gamma}$" is introduced for matching 
with experimental results. The appearence of the fine structure constant in this
treatment will be discussed in the last section.  
   	  
\section{Weak Static Limits of the New Equation}

The path equation (3.5) can be used as the equation of motion for any field
theory, constructed using the AP-geometry, provided that the theory has good
Newotonian limits. In such theories, (e.g. Mikhail and Wanas(1977), M{\o}ller
(1978),Hayashi and Shirafuji(1979)), the tetrad vectors  $\h{i}_{\mu}$
are considered as field variables. So, if we write 
$$
\h{i}_{\mu} = \delta_{_i \mu} + \epsilon h_{_i \mu}          \eqno{(5.1)} 
$$   	  
where $\epsilon$ is a small parameter, $\delta_{_i \mu}$ is Kronecker delta 
and $h_{i \mu}$ reprsents deviations from flat space, then the weak field condition 
can be fulfilled by neglecting quantities of the second and higher orders in 
$\epsilon$ in the expanded field quantities. For a static field assumption, we are 
going to assume the  vanishing of time derivatives of the field variables. 
 The vector components $Z^\mu$ defined by (3.2) will have the values
$$
Z^1 \approx Z^2 \approx Z^3 \approx \varepsilon {\  }{\  }{\  }    , Z^0 \approx 
1 - \varepsilon                  \eqno{(5.2)}   
$$    
where $\varepsilon (\approx  \frac{v}{c})$  is a  parameter. If we want to add the
condition of slowly moving particle to the previous conditions we should
neglect quantities of second and higher orders of the parameter $\varepsilon$. Thus, in
expanding the quantities of the path equation (4.3) we are going to neglect
quantities of orders $\epsilon^2, {\ }\varepsilon^2,{\  } \epsilon \varepsilon$
and higher,and also time derivatives of the field variable are to be neglected.
To the first order of the parameters, the only
field quantities that will contribute to the path equation (4.3) are given
by 
$$
\Lambda_{00}^{.\ .\ i} = - \epsilon h_{ 0 0,  i}   ,{\  }{\  }{\  }{\  }  (i= 1,2,3) 
                                                     \eqno{(5.3)}          
$$                     
$$             	 
\{^{\ i}_{0{\ }0}\} = \frac{\epsilon}{2} Y_{00,i}   ,{\  }{\  }{\  }{\  }  (i= 1,2,3) 
                                                     \eqno{(5.4)}
$$	
where $Y_{\mu \nu}$ is defined by 
$$
g_{\mu \nu}  = \eta_{\mu \nu} + \epsilon {\ } Y_{\mu \nu} {\ }{\ }{\ }, 
$$	
$g_{\mu \nu}$ is given by (2.2) and $ \eta_{\mu \nu}$ is the Minkowski metric tensor . Substituting from (5.3),(5.4) into (4.3)
 we get after some manipulations :
$$
\frac{d^2x^i}{d{\tau}^2} = -\frac{1}{2}{\ }\epsilon{\ }(1 -\frac{n}{2}\alpha \gamma)
Y_{00,i}{\ }Z^0{\ }Z^0.
                                                         \eqno{(5.5)}
$$  		
In the present case, the metric of the Riemannian space,  associated to AP-space,
can be written in the form, 
$$
(\frac{d\tau}{dt})^2 = c^2{\ }(1 + \epsilon {\ }Y_{00}).     \eqno{(5.6)}
$$
Substituting from (5.6) into (5.5) we get after some manipulations:
$$
\frac{d^2x^i}{dt^2} = - \frac{c^2}{2}{\ }\epsilon{\ }(1 -\frac{n}{2}\alpha \gamma)
{\ }Y_{00,i}{\ }{\ }{\ }{\ }(i=1,2,3)           
$$
which can be written in the form,
$$
\frac{d^2x^i}{dt^2} = - \pder{\Phi_s}{x^i} {\ }{\ }{\ }{\ }(i=1,2,3) {\ }{\ },
                                                   \eqno{(5.7)}
$$
where  
$$
\Phi_s =  \frac{c^2}{2}{\ }\epsilon{\ }(1 -\frac{n}{2}\alpha \gamma)
{\ }Y_{00}.                                                  \eqno{(5.8)}
$$	
Equantion (5.7) has the same form as Newton's equation of motion of 
a particle in a gravitational  field having the potential $\Phi_s$ given by 
(5.8), which differs from the classical Newtonian potential.
 In the case of macroscopic  particles $(n=0)$, we get from (5.8):
$$
\Phi_s =  \frac{c^2}{2}{\ }\epsilon{\ }Y_{00} = \Phi_N               \eqno{(5.9)} 
$$  	
where $\Phi_N$ is the Newtonian gravitational potential obtained from a similar
treatment of the geodesic equation. Thus (5.8) can be written in the form 
$$
\Phi_s =  (1 - {\frac{n}{2}} \alpha \gamma) \Phi_N.                  \eqno{(5.10)}
$$ 	
This last expression shows that the gravitational potential felt by the spinning 
particle is less than that felt by a spinless particle or a macroscopic test
particle. In other words, the Newtonian potentional is reduced for spinning 
particles, by a factor $(1 - {\frac{n}{2}} \alpha \gamma)$.  
    	
\section{Summmary and Discussion}	

The marriage between the two philisophical ideas of the present century, 
quantization and geometrization, has never been successful so far.  It is 
well known that quantization is successful in the domain of microphysics,
while geometrization is successful in the domain of macrophysics,
astrophysics and cosmology.  For example, the dynamics of microscopic particles is well described within the framework
of quantization, while the dynamics of macrophysical systems is described
successfully in the framework of geometrization.  The question now is: what
is the best description of the dynamics of microscpic particle in a background
gravitational field? Several authors have tried to answer this question, as
mentioned in section 1.  The problem is that their trails neither represent quantization of
geomerty, nor represent geometrization of quantum mechanics.

A solution of this problem may be, either starting by a certain geometry and using an 
appropriate procedure to see whether we can discover quantum features in this geometry; or 
starting in the quantum domain and using an appropriate procedure to look for 
geometric features.  We believe that any appropriate geometry describing nature
should contain paths that are quantized naturally, i.e.without using any 
quantization scheme, but how to discover such paths?  The trial given in the 
present work represents a step in this direction.  Although this trial is still
far beyond quantization of geometry, it gives a strong evidence that paths in the
AP-geometry are, in some sense, quantized.  We believe that this result is 
valid for any non-symmetric geometry (geometry with torsion), but this statement
needs confirmation.  Furthermore one can consider the present work as a step
twards unification of the dynamics of microscopic and macroscopic particles, since
the new path equation (3.5) could be applied for, macroscopic or microscopic,
massive or massless, and for spinning or spinless particles.

AP-spaces are defined as spaces whose structure, in four dimensions, is defined 
completely by a tetrad vector field subject to the condition (2.3) (cf.Robertson
(1932)).  There is no complete agreement between authors on whether the AP-spaces
are, in general, curved or flat.  Because of (2.15) many authors believe that
these spaces are flat (cf.Hayashi and Shirafuji (1979)).  Others believe that
these spaces are curved (cf.Mikhail and Wanas (1977)).  The cause is that for a
non-symmetric geometry (with a non-symmetric affine connexion) the curvature is
not uniquely defined.  However, all curvature tensors, in this space, are defined
in terms of the torsion tensor, and consequently the vanishing of the torsion will
leed to a flat space (Wanas and Melek(1995)). It is clear that the present work
solves this controversal problem . The geometric structure establised in the present 
work has the following features :\\
(1)  The structure of the space is defines completely (in 4-dimensions) by using
a tetrad subject to the condition (2.3) and thus, by definition, we are still
using an AP-geometry. So all what is done, from the gemetric point of view, is
that some hidden structures in this geometry are illuminated. We are goining to
call the structure developed in section 2 the Parametric Absolute Parallelism
(PAP)-Space. \\
(2) It is an affinely connected space endowed by a general non-symmetric connexion
(2.5). Thus the space possesses sufficient structure to carry out the operations of 
tensor analysis. \\
(3) At any point of the PAP- space we can define a metric tensor (2.2) which can be used
to carry out the operations of raising and lowering indicies.\\
(4) The space is certainly curved since the curvature tensors (2.9) and (2.11)
, corresponding to the non-symmetric connexion (2.5), are all
non-vanishing tensors. It is to be considered that other curvature tensors could
 be defined in the PAP-space by generalizing the AP-derivatives. All these 
 tensors are non vanishing and reduce to Riemann-Christoffel curvature tensor,
 of the Riemannian geometry, as b= 0. \\
(5) Paths in this geometry, corresponding to the connexion (2.5), are characterized
by equation (3.5), which is considered as a generalization of the geodesic equation 
of Riemannian geometry. \\
(6) The new structure covers both the Riemannian geometry (a = 1, b = 0) and the
conventional AP-geometry (a = 0, b = 1). Thus it is more general than both geometries.
The new structure can be used to solve the problem (raised by Wanas and Melek(1995)) of
constructing field theories, in non-symmetric geometries, that may be reduced to
GR for $b=0$ without any need  for vanishing torsion.

The torsion of the space-time is assumed to be generated by rotating sources. 
Many authors believe in this statement (cf. Ross (1989)) . We beleive that
torsion and metric tensors are both generic features of the gravitational field whether or not 
its source is rotating. Calculations show that, a sphecially symmetric solution
for a version of general relativity written in AP-space, torsion has non-vanishing 
components while the source of the field is non-rotating (cf. Wanas (1990)).

It is widely accepted that scalar particles (or macroscopic objects) cannot feel the torsion.
As stated by some authors, scalar test particles detect the metric of the space while
rotating test particles detect the torsion (cf. Nitch and Hehl(1980)).  This is similar to the situation
that neutral particles cannot feel the existence of  an electromagnetic field.
The question now is what is the intrinsic property of the
test particle that interacts with the background field ? In case if background
electromagnetic field the property is the electric charge. Similarly  if the background 
field is gravitational then the intrinsic property is the mass-energy of the particle
which interacts with the metric and/or the spin of the particle which interacts
with the torsion. 

	It has been shown (Wanas et al.(1995)b) that the results of applying the new
path equation (3.5), in the solar system, are consistent with the observational 
bases of GR.

     The author would like to thank Dr.M.Melek for many discussions, and Mr.M.E. 
Kahil for checking relevant calculations.

\section{References}
Bazanaski, S. L. (1977) Ann. Inst. H. Poincar\'{e},A {\bf{27}}, 145. 

~~~~~~~~~~~~~~~~~~(1989) J. Math. Phys. {\bf{30}}, 1018. \\
DeWitt, B. S. (1957) Rev. Modern Phys. {\bf{29}}, 377. \\
Galvao, C. and Teitelboim, C. (1980) J. Math. Phys. {\bf{21}}, 1863. \\
Hayashyi, K. and Shirafuji, T. (1979) Phys. Rev. D {\bf {19}}, 3524. \\
Hehl, F. W. (1971) Phys. Lett. {\bf{36}}A, 225. \\
Levi-Civita, T. (1950) "{\it {A Simplified Presentation of Einstein's United Field 
Equations}}",

~~~~~~~~~~~~~~~~~~~~~~~~~~~~~~~~~~~~~~~~~~~~~~~ translated pamphlet,  Blackie. \\         
Longo, M. J. (1988) Phys. Rev. Lett. {\bf{60}}, 173. \\
Melek. M. (1988) Acta Physica Solvaca {\bf{38}},  146. \\ 
Mikhail, F. I. (1952) Ph.D. Thesis, University of London.\\      
Mikhail, F. I. and Wanas, M. I. (1977) Proc. Roy. soc. Lond. A {\bf{356}}, 471.\\
M{\o}ller, C. (1978) Mat. Fys. Medd. Dan. Vid. Selek. {\bf{39}},1. \\
Nitch, J. and Hehl, F. W. (1980) Phys. Lett. {\bf{90}}B, 98. \\
Robertson, H. P. (1932) Ann. Math. Princeton {\bf{33}}, 493. \\                           
Ross, D. K. (1989) Int. J. Theort. Phys. {\bf{28}}, 1333. \\
Wanas, M. I. (1990) Astron. Nachr. {\bf{ 311}}, 253. \\ 
Wanas M. I. and Melek M. (1995) Astrophys. Space Sci, {\bf{228}}, 277. \\
Wanas, M. I., Melek, M. and Kahil, M. E. (1995)a Astrophys. Space Sci,{\bf{228}}, 273. 
 
~~~~~~~~~~~~~~~~~~~~~~~~~~~~~~~~~~~~~~~~~~~~~~~~~~~~(1995)b to be submitted for publication.     
\end{document}